\documentclass[doublecol,figures]{epl2} 
\usepackage[latin1]{inputenc}
\usepackage{color}
\title{Sampling bias in systems with structural heterogeneity and limited internal diffusion}
\shorttitle{Sampling bias}
\author{J.-P. Onnela\inst{1,2,3}, N. F. Johnson\inst{4}, S. Gourley\inst{1,2}, G. Reinert\inst{5}, \and M. Spagat\inst{6}}
\shortauthor{J.-P. Onnela \etal}
\institute{                    
  \inst{1} Department of Physics, University of Oxford, Oxford OX1 3PU, UK\\
  \inst{2} CABDyN Research Cluster, Said Business School, University of Oxford, Oxford, OX1 1HP, UK\\
  \inst{3} DBEC, Helsinki University of Technology, P.O. Box 9203, FIN-02015 HUT, Finland\\
  \inst{4} Physics Department, University of Miami, Coral Gables, Florida 33124, USA\\
  \inst{5} Department of Statistics, University of Oxford, Oxford OX1 3TG, UK\\
  \inst{6} Department of Economics, Royal Holloway, University of London, TW20 0EX, UK
}


\abstract{Complex systems research is becomingly increasingly data-driven, particularly in the social and biological domains. Many of the systems from which sample data are collected feature structural heterogeneity at the mesoscopic scale (i.e. communities) and limited inter-community diffusion. Here we show that the interplay between these two features can yield a significant bias in the global characteristics inferred from the data. We present a general framework to quantify this bias, and derive an explicit corrective factor for a wide class of systems. Applying our analysis to a recent high-profile survey of conflict mortality in Iraq suggests a significant overestimate of deaths.}

\begin{document}
\maketitle

\section{Introduction}
Monitoring large social or biological systems bears similar challenges to monitoring many-particle systems in physics. The increasing availability of data on human behaviour from information and communication technologies \cite{jppnas,marta} and data from high throughput techniques in biology enable scientists to study these diverse systems with similar methodologies. Many biological and social systems are not internally homogeneous, but instead feature time-dependent community groupings and limited inter-community mixing \cite{jppnas,marta,palla,schnell}. Individuals form dynamic groups in professional and private settings reflected in, for example, structures of scientific collaboration and mobile phone call patterns \cite{palla}. The cell nucleus consists of multiple compartments with different micro-environments that exist in spatially localised regions in the heterogeneous intranuclear space \cite{schnell}. This problem setup is similar to that in so-called metapopulation models, which involve spatially structured populations and are commonly used in ecology and epidemiology \cite{hanski:1998,moilanen:2002}. In this Letter, we quantify the consequences of sampling a subset of objects in such a system. Starting with a general theoretical framework, we show that the interplay between heterogeneity and limited diffusion can yield a substantial bias in the inferred global characteristics. We obtain an explicit corrective factor to offset a bias that occurs if the structural heterogeneity of the system and the limited internal diffusion within the system are not taken into account in the initial data sampling. We then consider a special case of this general framework, in which the corrective factor turns out to only depend on three parameters. Two of these parameters are associated with heterogeneity and one with diffusion. Finally, we consider the specific example of a recent conflict mortality study in Iraq, and show that a considerable positive bias likely arose in the inferred mortality numbers.

\section{General framework}
Consider a large system made up of $N$ particles characterised by a microscopic state variable $x_{i}$. The system is heterogeneous in that it consists of $m$ different subsystems or communities $S_{1}, \ldots, S_{m}$ with $N_{i}$ particles in $S_{i}$ such that $N_{1} + \ldots + N_{m} = N$. The subsystems are interconnected in some limited way, thereby allowing for only partial diffusion or mixing of particles between them.  We wish to learn about the state of the system described by the extensive macroscopic variable $X=\sum_{i=1}^{N} x_{i}$ but, in line with typical empirical scenarios, assume that we cannot observe the entire system. Instead, we monitor the state of a set of tagged particles in different subsystems and use this data to make statistical inferences about $X$.

\begin{figure}[]
\includegraphics[width=1\linewidth]{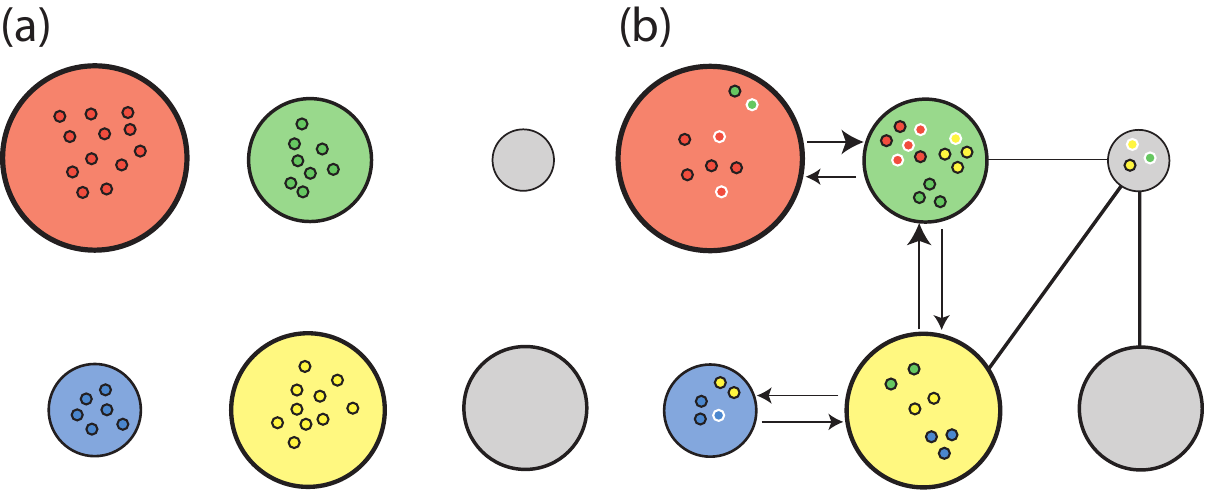}
\caption{(Colour online) (a) The system is prepared by tagging some particles in some of the subsystems, which corresponds to a sampling process. The particles are non-interacting and indistinguishable apart from the initial subsystem given by their colour. (b) After the initial state, the matrix $\mathbf{f} \ne \mathbf{I}$ quantifies mixing between subsystems. It can be interpreted as a weighted and directed network adjacency matrix of the subsystems. The state of particle $x_i \in \{0,1 \}$ is indicated by colouring its circumference black or white, respectively. Only tagged particles are visible and available for analysis.}
\label{fig:schem}
\end{figure}

Let us assume that the particles are identical and non-interacting and that each can be in 
one of two states $x_{i} \in \{0,1 \}$. The system is initially prepared with $x_{i}=0 $ for all $i$ and 
only irreversible $0 \to 1$ changes are considered. Microscopic state changes are subsystem specific, with the element $q_{k}$ of the vector $\vec{q}$ specifying the probability for a particle in subsystem $S_k$ to change state. Hence the $x_{i}$ are independent and identically distributed random variables within a given subsystem and we let $y_k$ denote a random variable having the distribution of any $x_{i}$ present in $S_{k}$. The state of a particle can be identified with, for example, the staining of cancerous cells in a biological organism under medical imaging (stained vs. clear), or the disease status of an individual (healthy vs. diseased). The subsystem specific probabilities $\vec{q}$ could arise from there being different numbers of cancerous cells or pathogens in these systems. 

The mixing of particles is governed by the constant mixing matrix $\mathbf{f} = [\vec{f_{1}} \vec{f_{2}} \cdots \vec{f_{m}}]$, where $\vec{f_{i}}$ specifies the fraction of time particles initially placed in $S_{i}$ spend in other subsystems. The entries of $\mathbf{f}$ can be interpreted as probabilities of finding particles in different subsystems (see Fig.~1). The diagonal elements $f_{ii}$ correspond to the probability of finding a particle in its initial subsystem. Note that $\mathbf{f}$ does not need to be symmetric. In the limit as the mobility of the particles tends to zero, the matrix $\mathbf{f}$ consists of only diagonal elements $f_{ij} = \delta_{ij}$, with the effect that the subsystems become completely isolated.

Denote by $X_{i}$ the contribution of all particles initially in $S_{i}$ towards $X$, and by  $X_{ij}$  the contribution of a single particle $j=1, \ldots, N_{i}$ initially in $S_{i}$ towards $X_{i}$. Let $D_{ik}$ denote the number of particles initially in $S_i$ which are observed in $S_k$; then $D_{ik}$ follows a binomial distribution with parameters $N_i$ and $f_{ik}$. We write our quantity of interest $X$ as 
\begin{equation}
X = \sum_{i=1}^{m} X_{i} = \sum_{i=1}^{m} \sum_{j=1}^{N_{i}} X_{ij}.
\label{eq:mobsum}
\end{equation}
Consider now a situation where, to estimate $X$, we can draw samples from only some of the subsystems. Let $S$ consisting of $m$ subsets $S_{k}$ denote the set of all subsystems and let $S' = \bigcup_{k=1}^{m'} S_{k}$, i.e. the first $m'$ of these sets, denote the set of samplable subsystems. The expectation value of $X$ in the entire system, $\langle X \rangle_{S}$, and in the samplable system, $\langle X \rangle_{S'}$, is given by
\begin{eqnarray}
\langle X \rangle_{S}  & = & \sum_{i=1}^{m} \langle X_{i} \rangle =
\sum_{i=1}^{m} \sum_{k=1}^m \langle D_{i k} \rangle \langle y_k \rangle =
 \sum_{i=1}^{m} N_{i} \sum_{k=1}^{m} f_{ik} \langle y_k \rangle \nonumber \\
\langle X \rangle_{S'}  & = & \sum_{i=1}^{m'} \langle X_{i} \rangle = \sum_{i=1}^{m'} N_{i} \sum_{k=1}^{m} f_{ik} \langle y_k \rangle, 
\end{eqnarray}
respectively. If the subsystems are heterogeneous and this is not accounted for in the sampling procedure, we may incur a significant bias. To quantify this, we define the \emph{bias factor} $R$ as the scaled ratio of $\langle X \rangle_{S'}$ to $\langle X \rangle_{S}$, 

\begin{equation}
R = \frac{\sum_{i=1}^{m'} N_i  \sum_{k=1}^{m} f_{ik} \langle y_k \rangle /N'} {\sum_{i=1}^{m} N_{i} \sum_{k=1}^{m} f_{ik} \langle y_k \rangle / N},
\label{eq:r}
\end{equation}
where $N' \le N$ is the number of particles in $S'$ and values of $R>1$ ($R<1$) correspond to overestimating (underestimating) 
the expectation value of $X$ in the system when sampling is based on subsystems in $S'$ only. 

\section{Specialised framework}
A special case of the framework arises when  the microscopic state variables $x_{i}$ and $y_k$ correspond to independent Bernoulli trials related to some event $\omega$. We assume that the event $\omega$ occurs independently of the mixing. Now $q_{i}$ ($1-q_{i}$) is the probability of observing $x=1$ ($x=0$) in subsystem $S_i$ long enough after the initial state so that the system has reached an equilibrium. Regardless of the number of subsystems present, the system can always be divided into a samplable subsystem and a  non-samplable subsystem. Let $S_{I} = S'$ and let the remaining subsystems form the non-samplable subsystem $S_{O} = \bigcup_{k=m'+1}^{m} S_k$. As a mnemonic, the subscript $I$ refers to in-sample and $O$ to out-of-sample. Note that whereas before $S' \subseteq S$, here $S_{I} \cap S_{O} = \emptyset$, so that although $S_{I} = S'$, $S_{O} \ne S$. We now have $N_{I} = N' = \sum_{k=1}^{m'} N_{k}$ and $N_{O} = \sum_{k=m' + 1}^{m} N_{k}$, corresponding to the number of particles in $S_{I}$ and $S_{O}$, respectively, and $N_{I} + N_{O} = N$. We define the 'renormalised' probabilities $q_{I} = N_{I}^{-1} \sum_{k=1}^{m'} N_{k} \, q_{k}$ for a particle to be subjected to $\omega$ while present in $S_{I}$ and its complement $1-q_{I}$ for the particle to not be subjected to $\omega$ while present in $S_{I}$. Similarly, we define for $S_{O}$ the probability $q_{O} = N_{O}^{-1} \sum_{k=m' + 1}^{m} N_{k} \, q_{k}$ (and its complement $1-q_{O}$) for a particle to (not) be subjected to $\omega$ while present within $S_{O}$. Finally, we define the mobility factors such that $f_{I}$ ($f_{O}$) is the probability for a particle initially placed in $S_{I}$ ($S_{O}$) to be present within $S_{I}$ ($S_{O}$), and $1-f_{I}$ ($1-f_{O}$) is the probability for a particle initially placed in $S_{I}$ ($S_{O}$) to not be present within $S_{I}$ ($S_{O}$), i.e., to be present within $S_{O}$ ($S_{I}$). These are written as 

\begin{eqnarray}
f_{I} & = & N_{I}^{-1} \sum_{i=1}^{m'} N_{i} \sum_{j=1}^{m'} f_{ij} \nonumber \\
f_{O} & = & N_{O}^{-1} \sum_{i=m'+1}^{m} N_{i} \sum_{j=m'+1}^{m} f_{ij}.
\end{eqnarray}

We now define $\pi_{\alpha \beta }$ with $\alpha, \beta \in \{O,I\}$ as the probability that a particle picked uniformly at random was 
placed initially in $S_{\alpha}$ 
with $x_{i}=0$ and changes state to $x_{i}=1$ in $S_{\beta}$. This leads 
to $\pi_{OO} = \frac{N_{O}}{N_{I}+N_{O}} \, f_{O} \, q_{O}$, $\pi_{OI} = \frac{N_{O}}{N_{I}+N_{O}} \, (1-f_{O}) \, q_{I}$, 
$\pi_{IO} = \frac{N_{I}}{N_{I}+N_{O}} \, (1-f_{I}) \, q_{O}$, and $\pi_{II} = \frac{N_{I}}{N_{I}+N_{O}} \, f_{I} \, q_{I}$. 
The sum $\pi_{II} + \pi_{IO} + \pi_{OI} + \pi_{OO}$ is the probability that a randomly chosen particle is subjected to 
$\omega$ and hence changes its microscopic state. The expected number of particles with $x_{i}=1$ in a population of size $N$ is hence 
$  N_{O} \, f_{O} \, q_{O} 
+ N_{O} \, (1-f_{O}) \, q_{I} + \, N_{I}  \, (1-f_{I}) \, q_{O} + N_{I} \, f_{I} \, q_{I}
=(q_I-q_O)(f_IN_I-f_ON_O) + q_IN_O + q_ON_I$, whereas the probability that a randomly chosen particle in $S_{I}$ changes state is 
$q_{I}f_{I} + q_{O}(1-f_{I})$. Hence the expected number of realisations for a population of size $N$, based on the rate for $S_{I}$ only, 
would be $(N_I+N_O) [q_If_I + q_O(1-f_I)]$. We obtain 

\begin{equation}
R = \frac{(N_I+N_O)[q_If_I + q_O(1-f_I)]}{(q_I-q_O)(f_IN_I-f_ON_O) + q_IN_O + q_ON_I}.
\end{equation}
Assuming that $N_{I} \ne 0$ and $q_{O} \ne 0$, and setting $q = q_{I} / q_{O}$ and $n = N_{O}/N_{I}$, we obtain

\begin{equation}
R = R(f_I, f_O, q, n) = \frac{(1+n)(1+qf_I - f_I)}{(q-1)(f_I-f_On)+qn+1}.
\label{eq:r}
\end{equation}
Hence the bias factor $R$ depends only on $f_{I}$, $f_{O}$, and the ratios $q = q_{I} / q_{O}$ and $n = N_{O}/N_{I}$. Finally, in the case of symmetric mobility with $f_{I}=f_{O}=f$, the above expression simplifies to 

\begin{equation}
R = R(f,q,n) = \frac{(1+n)(1+qf-f)}{f(q-1)(1-n)+qn+1}.
\label{eq:specialr}
\end{equation}

The no-bias limit of $R=1$ requires either (1) $n=0$ (i.e. $N_{O}=0$) implying that no particle is placed initially in $S_{O}$, or (2) $q=1$ (i.e. $q_{I} =q_{O}$) implying equal rates of changing state in $S_{I}$ and $S_{O}$, or (3) $f =1/2$ which suggests that particles based in $S_{I}$ spend on average half of their time in $S_{O}$ and vice versa. Setting $R(f,q,n)=r$ for general $r$ and solving for $q$ in terms of $n$ and $f$ yields 

\begin{equation}
q(f,n,r) = \frac{f(1 + n + n r - r) + r - n - 1}{f(1 + n + nr - r) - nr}.
\end{equation}

Although $q$ is unobservable, we can estimate ${\tilde q} = N^{-1}  \sum_{i,j} X_{ij}$ and 
${\tilde q'} = (N')^{-1}  \sum_{i=1}^{m'} \sum_{j=1}^{N_i} X_{ij}$, leading to the asymptotically unbiased estimator $\hat{R} = \tilde{q}' / \tilde{q}$ for the bias factor $R$. If $R=1$ then we would expect that $\hat{R} \approx 1$. The variation in $\hat{R}$ can be assessed via a normal approximation \cite{book}. Basing $\tilde{q}'$ on $S_I$ and assuming that $\langle X \rangle_S $ is not too small, the approximate variance is 

\begin{eqnarray}
\textrm{Var}(\hat{R}) \approx \frac{(1+n)^2}{q_0 N_I\big(f(q-1)(1-n) + qn + 1\big)^2} \times \nonumber \\
 \big(fq(1-q_I) + (1-f) (1- q_O) + f(1-f)(q q_I + q_O - q) \big).
\end{eqnarray}

\section{Application to conflict mortality}
We will now exemplify the above framework by applying it to study conflict mortality. To estimate the number of deaths in a conflict, one would ideally like to have access to a complete national list of households from which a sample could be drawn at random. Even when this scenario is feasible, the selected households are widely scattered, which is costly not only in terms of time and money, but also exposes the researchers to high levels of risk. To overcome these concerns, recent studies economise resources by using a cluster sampling methodology. This hierarchical sampling process involves making choices on how to choose large geographic areas and how to proceed from them to individual households.

We can equate particles in the framework with individuals such that the system size $N$ corresponds to the population of the country 
and the state of each particle $x_{i} \in \{0,1 \}$ corresponds to the individual being alive or dead (where the death has resulted from 
conflict related violence), respectively. The different subsystems correspond to heterogeneous areas that are characterised by varying levels of conflict related violence such that the probability for an individual to be killed when he or she is in $S_{k}$ is given by $q_{k}$ regardless of where his or her residence is located. Note that these areas, or zones, may be fragmented and inter-dispersed. Now $\langle X_{k} \rangle$ corresponds to the expected number of deaths in $S_{k}$ for a given $q_{k}$, and $\langle X \rangle$ corresponds to the expected number of deaths in the country. Daily human movement between different areas is quantified by the mixing matrix. The initial subsystem of a particle can be identified with the residential zone of the individual. The 'renormalised' systems $S_{I}$ and $S_{O}$ correspond to sets of subsystems that may or may not be sampled, respectively, given the sampling method. To include an individual in the study, his or her home needs to be located in the samplable subsystem $S_{I}$. 

Let us consider a situation in which data has been collected using a sampling procedure and we are concerned that this sampling procedure may not be sufficiently sensitive to the structural heterogeneity of the system and the limited internal diffusion within the system, so that a systematic bias may arise. We can then use the proposed framework, after the initial data collection, to offset the bias resulting from not having taken these factors fully into account.

\begin{figure}
\begin{center}
\includegraphics[width=1.0\linewidth]{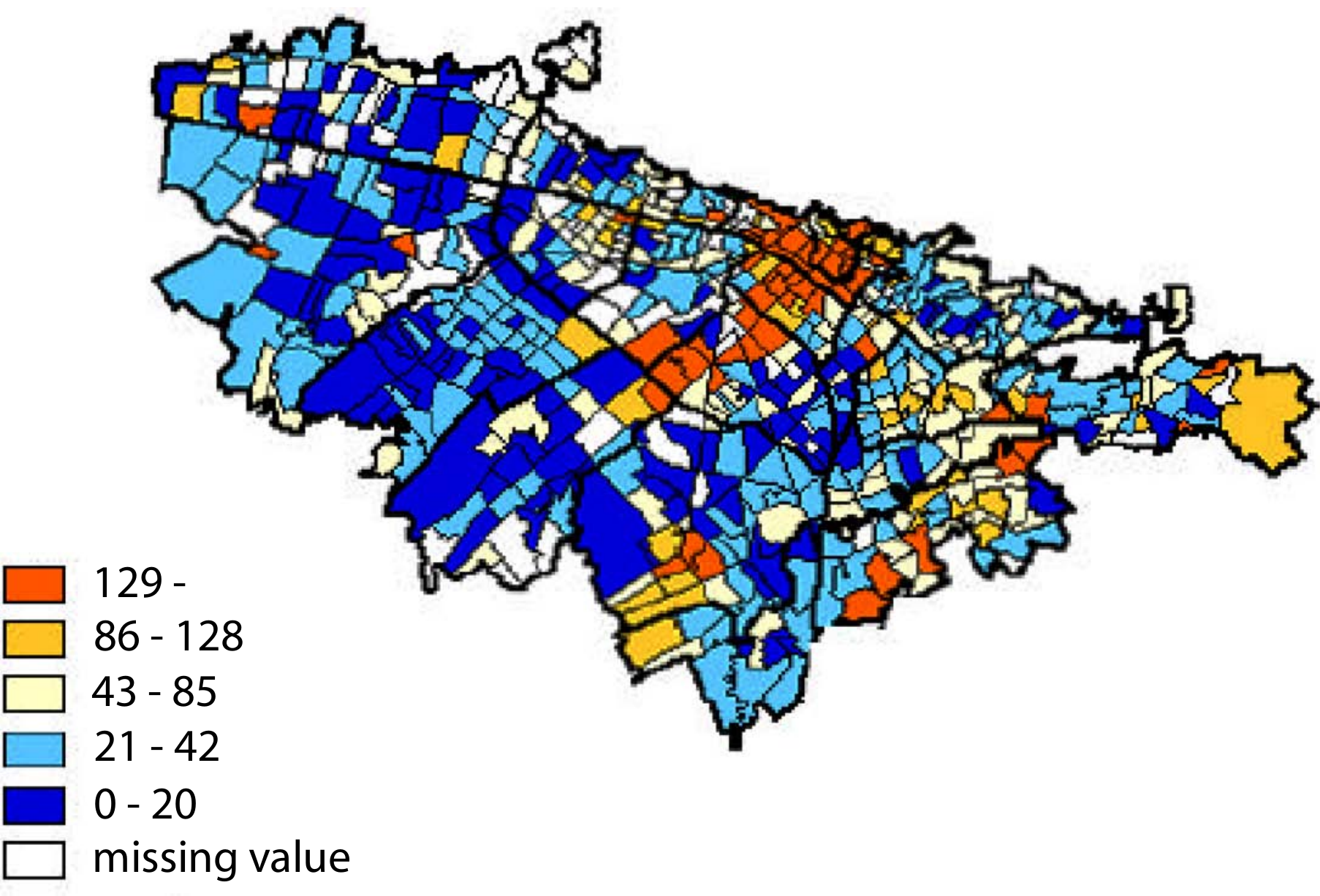}
\caption{(Colour online) Structural heterogeneity in a social system under conflict. The map shows the average homicide rate according to censual sectors in Bogot\'a, Colombia, in the period 1997-1999. Source: Instituto Nacional de Medicina Legal. Figure adapted from \cite{bogota}.}
\label{fig:bogota}
\end{center}
\end{figure}

Structural heterogeneity between subsystems in the context of conflict mortality is exemplified in Fig. \ref{fig:bogota}, which shows how urban violence varies from neighbourhood to neighbourhood, in this case in Bogot\'a, Colombia. Similar patterns hold for cities worldwide \cite{ourwebsite}. While these data are based mostly on criminal violence as opposed to conflict violence, it is plausible that, similarly, a spatially inhomogeneous pattern holds for conflict violence. Each cell in the map can be associated with one of the $k=1,\ldots,m$ subsystems and the colouring reflects a realised value of $X_{k}$. In a completely homogeneous system with $q_{1}=\cdots=q_{m}$ we have $\langle X_{1} \rangle = \cdots = \langle X_{m} \rangle$ and would expect to observe less fluctuation in the values of $X_{k}$. We conjecture that structural heterogeneity is likely to hold in conflict areas.

\begin{figure}
\begin{center}
\includegraphics[width=0.7\linewidth]{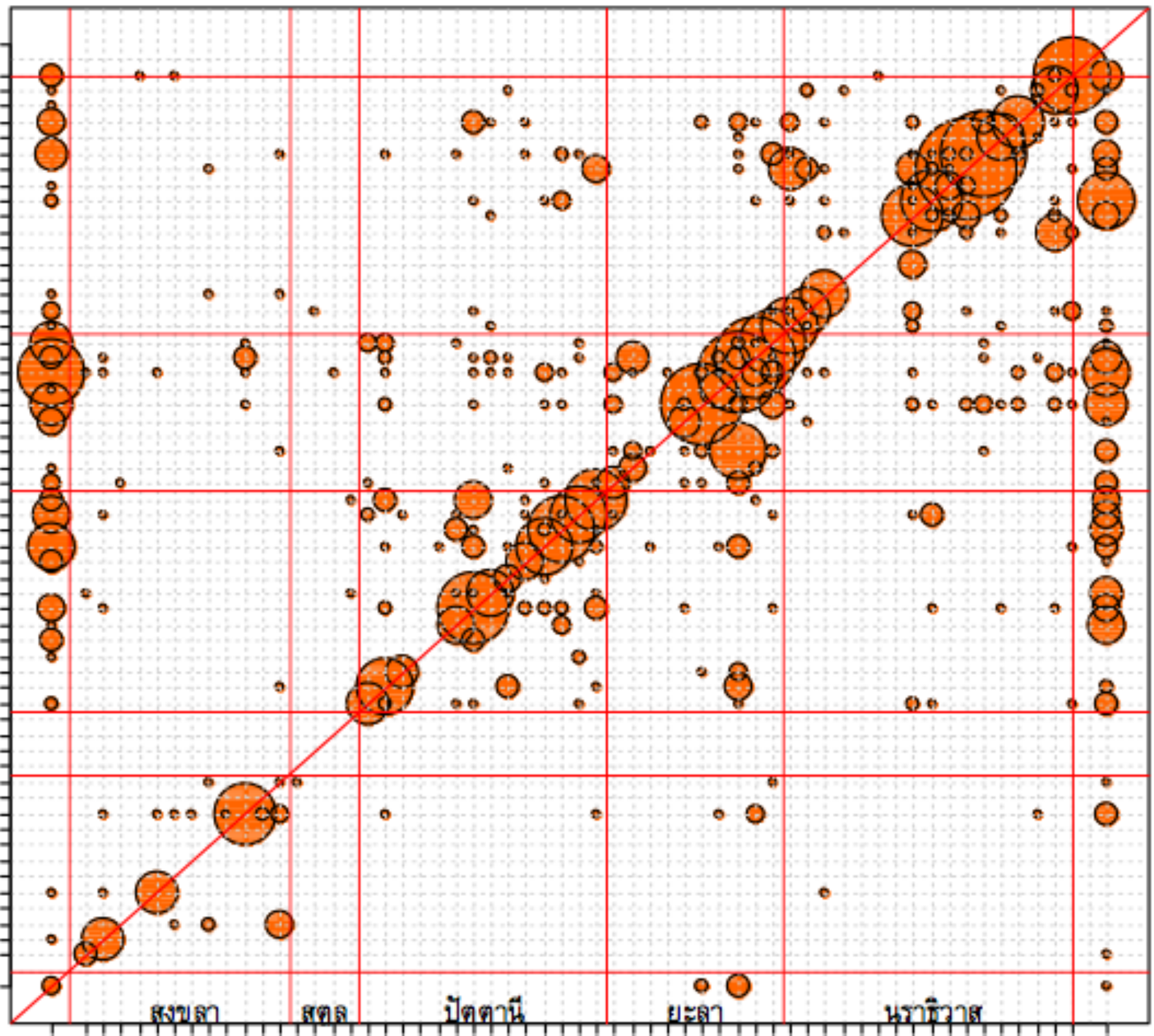}
\caption{(Colour online) Limited internal diffusion in a social system under conflict. The relationship between the residence of casualties (killings and injuries) and the place where they were attacked. The axes correspond to 59 distinct spatial locations listed in identical order, such that the horizontal axis represents the residence of the  casualties while the vertical axis represents the place where the incident occurred. The data are from a conflict in Thailand and they are based on a hospital monitoring system. The bubble plots reflect the number of casualties in each area. Figure adapted from \cite{thaistudy}.}
\label{fig:thai}
\end{center}
\end{figure}

Limited internal diffusion between subsystems in the context of conflict mortality is exemplified in Fig. \ref{fig:thai}, which shows the location of residence of the victims (horizontal axis) and the location of attacks (vertical axis) in a conflict in Thailand. This matrix can be interpreted to reflect the underlying diffusion matrix $\mathbf{f}$ and it is useful to consider two limiting cases. First, if the matrix were completely scattered, there would be no correlation between the location of residence and the location of violence. In this case the choice of sampling locations and the locations of violence are  uncorrelated, and one might wish to choose sampling locations that are easily accessible. These sampling locations might be inherently more or less violent than the system at large but, due to extensive mobility of individuals, the choice of sampling locations would not induce a systematic bias. Second, if the matrix were perfectly diagonal, there would be a one-to-one correlation between the location of residence and the location of violence. If the sampling locations were, say, more violent than the system at large, due to lack of mobility between subsystems, the overall estimate would be biased upward. In both scenarios one would need to take population densities into account. We conjecture that diffusion between subsystems is very limited under conflict.

\begin{figure}
\begin{center}
\includegraphics[width=1\linewidth]{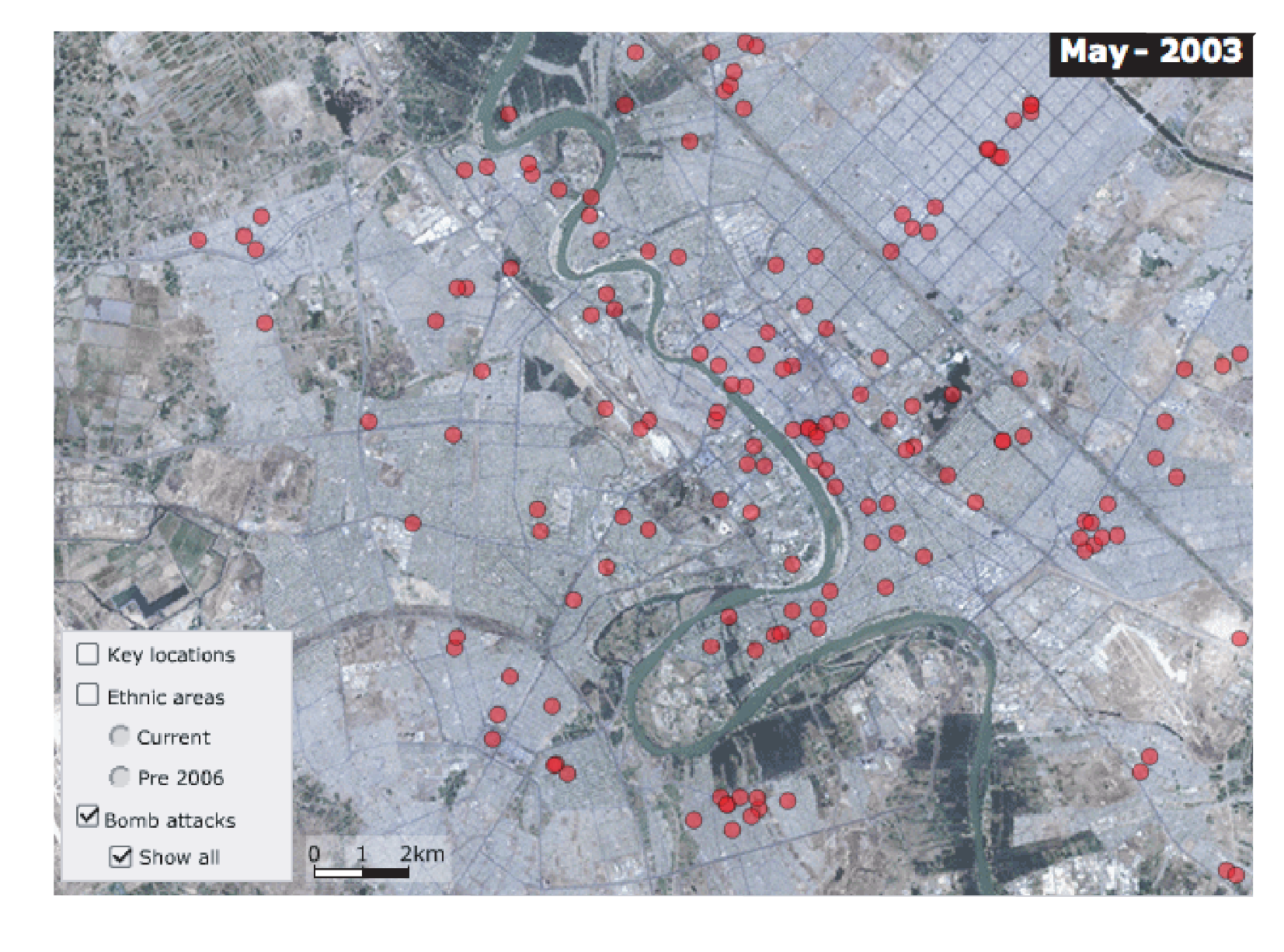}
\caption{(Colour online) A satellite image of Baghdad showing the position of attacks that resulted in more than 10 dead. ``[The attacks] are located as accurately as possible from reports since 2003. Where an exact location is not possible, in areas such as Sadr City, the marker has been placed within the district.''  The locations of the attacks coincide with the structure of underlying road network with most attacks taking place either on major roads or on roads off major roads. Image and quotation adapted from BBC News\cite{bbc}.}
\label{fig:bbc}
\end{center}
\end{figure}

We now focus on the final stages of the sampling procedure that was used estimate conflict mortality in Iraq \cite{L2}, and refer to it as the \emph{Cross Street Sampling Algorithm (CSSA)}: (1) Select a ``constituent administrative unit'' proportionally to their estimated population size, (2) select a main street from ``a list of all main streets'', (3) select randomly a residential street from ``a list of residential streets crossing the main streets'', (4) enumerate the households on the street, (5) select one household at random to initiate the interviewing, proceeding to 39 further adjacent households. Fig.~\ref{fig:bbc} demonstrates that violent events  tend to be focused around cross-streets. Because cross-streets are chosen for sampling, the location of violence and the location of sampled sites are correlated by means of accessibility. This correlation results in a biased estimate of deaths and is further amplified due to minimal mixing of populations between the zones.

To apply the above framework we need values for the model parameters. The population parameter $n=N_{O}/N_{I}$ gives the proportion of population resident in $S_{O}$ to that resident in $S_{I}$. Street layouts in Iraq are often irregular, hence CSSA will miss any neighbourhood not in the immediate proximity of a cross-street. Analysis of Iraqi maps suggests $n=10$ is plausible \cite{msb}. The violence parameter $q=q_{I}/q_{O}$ gives the relative probability of death for anyone present in $S_{I}$, regardless of their zone of residence, to that of $S_{O}$. For conflicts like the one in Iraq, violent events tend to be focused around cross-streets since they are a natural habitat for patrols, convoys, police stations, parked cars, roadblocks, cafes and street-markets. Major highways would not offer such a wide range of potential targets -- nor would secluded neighborhoods and, therefore, the streets that define the samplable region $S_{I}$ are prime targets for improvised explosive devices, car bombs, sniper attacks, abductions and drive-by shootings \cite{msb}. Given the extent and frequency of attacks, $q=5$ is plausible \cite{msb}. The diffusion parameter $f=f_{I}=f_{O}$ gives the fraction of time spent by residents of $S_{I}$  ($S_{O}$) in $S_{I}$ ($S_{O}$). Given the nature of the violence, travel is limited; women, children and the elderly tend to stay close to home. Consequently, mixing of populations between the zones is minimal. Using the time people spend in their homes as a lower bound on the time they spend in their zones, assuming that there are two working-age males per average household of seven \cite{L2}, with each spending 6h per 24h day outside their own zone, yields $f=f_{I} = f_{O} = 5/7 + 2/7 \cdot 18/24=13/14$ \cite{msb}. These values yield $R=3.0$, suggesting that the Iraq estimate \cite{L2} provides a substantial overestimate of deaths.

\begin{figure*}
\begin{center}
\includegraphics[width=0.68\linewidth]{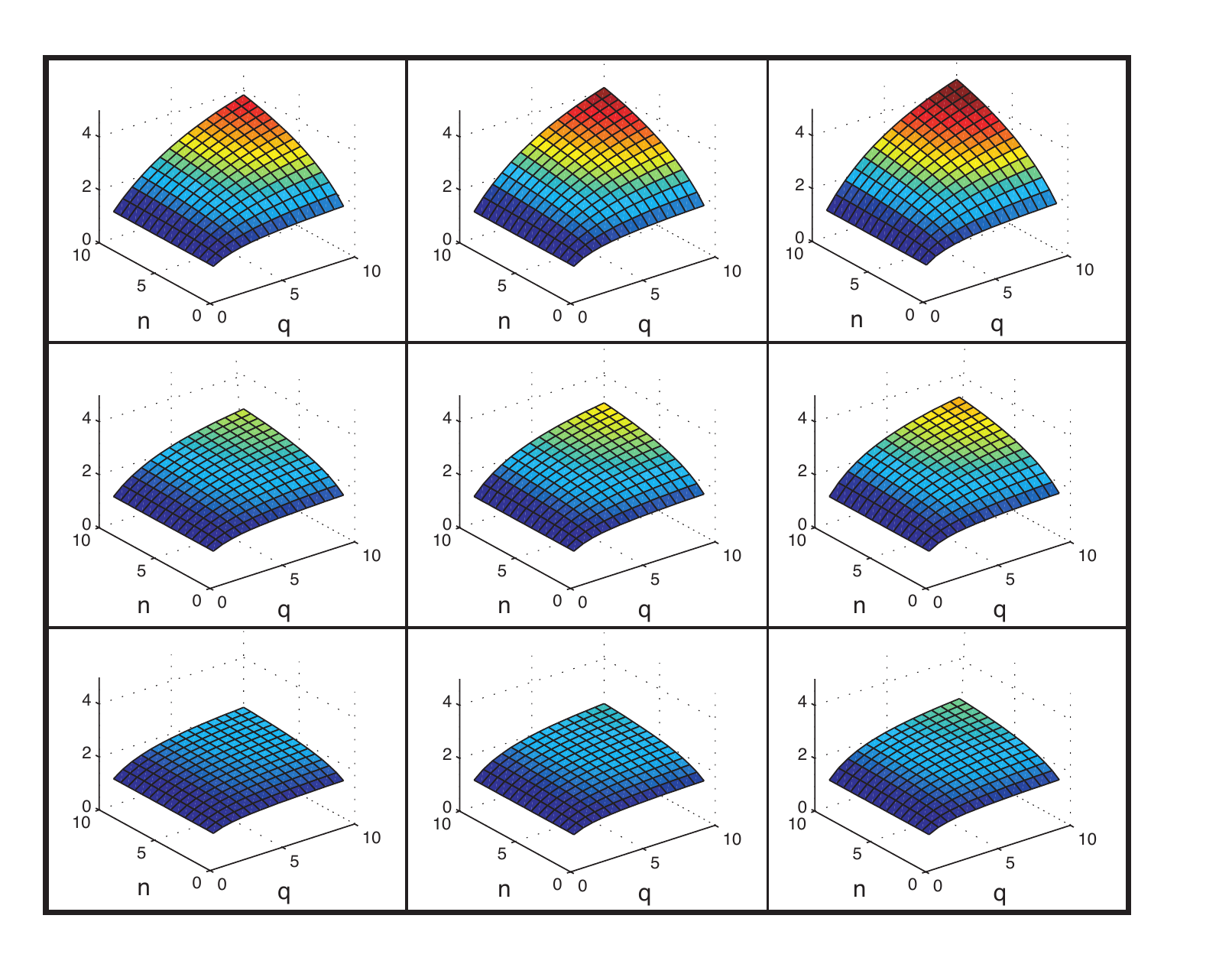}
\caption{(Colour online) Sensitivity analysis of bias factor $R$ defined by Eq.~\ref{eq:r}. Each panel shows $R=R(f_{I}, f_{O}, q, n)$ with the values of $f_{I}$ and $f_{O}$ fixed for each panel. Here $f_{I}$ ($f_{O}$) varies by columns (rows) over the values $\{0.75, 0.85, 0.95\}$ increasing from left to right (bottom to top). The height of the surface from the $(n,q)$ surface, in addition to being given by the $z$-coordinate in the plots, is also colour coded to guide the eye and to emphasise the smoothness of the surfaces.}
\label{fig:sens}
\end{center}
\end{figure*}

It is clear from Eq.~\ref{eq:r} that in order to arrive at an accurate estimate of $R$, one needs to have reasonably accurate estimates of the parameters $f_{I}, f_{O},q$ and $n$. To gauge the sensitivity of our result, we perform a simple sensitivity analysis by evaluating $R$ for different values of parameters in Fig.~\ref{fig:sens}. This shows the effect of relaxing the constraint $f=f_{I}=f_{O}$ and it is clear that in the limit of no mobility ($f_{I}=f_{O}=1)$ the bias is greatest. Conceptually speaking, the bias emerges from having simultaneously partial localisation of violence (structural heterogeneity) and partial localisation of people (limited internal diffusion). Both of these conditions are needed for the bias to emerge, since if $q=1$ (structural homogeneity) we have $R=1$ regardless of $n$ and $f$, and if $f =1/2$ (perfect diffusion), we have $R=1$ regardless of $q$ and $n$. In general, the shapes of the $R$-surfaces in Fig.~\ref{fig:sens} are smooth and the surfaces are monotonically increasing functions of $n$ and $q$. In this sense the framework is robust to the parameter values.

A more precise quantification of the bias can be achieved within the framework only if the actual micro-level data of the conflict study \cite{L2} are released, which would enable a more precise determination of the model parameters. Importantly, this does not entail further data collection, which is especially valuable when the survey needs to be carried out under extremely difficult conditions. Even release of information concerning how many streets are included in ``a list of all main streets'' in step (2) of CSSA would improve the estimate. This is because  the definition of a ``main street'' sets the granularity level of the system. A shorter list implies that the areas enclosed by the streets are bigger, which necessarily decreases mixing between areas, and results in an even larger bias.

\section{Conclusion}
We have presented a framework that can be used to gauge sampling bias in systems featuring heterogeneity and limited internal diffusion. We have applied the framework to a recent conflict mortality study \cite{L2} to illustrate how one can, after the initial data collection, adjust for the bias resulting in sampling such a system. We have demonstrated that the conflict mortality study is likely to present a high upward bias and, using our framework, have gauged the extent of this bias using simple plausibility arguments for our framework parameters. We believe that our approach and assumptions are reasonable given the limited information to hand. It appears that the results reported in \cite{L2} are a substantial overestimate of deaths. This finding is compatible with recent independent research. The figures reported in \cite{L2} are 3 times higher than the Iraq Living Conditions Survey of the UN Development Program estimate for the same time period (the first 13 months of the war) \cite{ILCS}, 4 times the Iraq Family Health Survey estimate for the same time period \cite{IFHS}, and 12 times the Iraq Body Count estimate (based on media monitoring) for the same time period \cite{IBC}. Rather than opening a debate on the precise extent of the bias, we hope that the present work will open up the way to further studies aimed at specifying more precisely the information needed to improve these estimates. Given that many social and biological systems feature structural heterogeneity and limited internal diffusion, our framework should prove invaluable for correcting for such biases.

\vspace{1cm}

\acknowledgments
JPO acknowledges Wolfson College, Oxford. GR acknowledges MMCOMNET, Grant No. FP6-2003-BEST-Path-012999. We acknowledge Ph.D. candidate Kraiyos Patrawart for his English translation of the report in reference \cite{thaistudy}.

\end{document}